\documentclass[aps,prd,twocolumn,superscriptaddress,showpacs]{revtex4}
\usepackage[dvips]{graphicx}
\usepackage{ulem}
\usepackage[usenames]{color}
\usepackage{amsmath}
\usepackage{float}

\begin{document}

%\preprint{}

\title{Nature of charmed strange baryons $\Xi_c(3055)$ and $\Xi_c(3080)$}
%\title{Newly reported charmed strange baryons in the $^3P_0$ model}


\author{Ze Zhao}
%\email{washingtonze@shu.edu.cn}
%\thanks{}
\affiliation{Department of Physics, Shanghai University, Shanghai 200444, China}

%\author{Ailin Zhang}
%\email{??????????}
%\thanks{}
%\affiliation{Department of Physics, Shanghai Uinversity, Shanghai 200444, China}
\author{Dan-Dan Ye}
%\email{??????????}
%\thanks{}
\affiliation{Department of Physics, Shanghai University, Shanghai 200444, China}
\affiliation{College of Mathematics, Physics and Information Engineering, Jiaxing University, Jiaxing 314001, China}

\author{Ailin Zhang}
\email{zhangal@staff.shu.edu.cn}
%\thanks{}
\affiliation{Department of Physics, Shanghai University, Shanghai 200444, China}
%\affiliation{Department of Physics, Shanghai Uinversity, Shanghai 200444, China}

%\date{July 13, 2016}
\begin{abstract}
The hadronic decay widths and some ratios of branching fractions of the newly observed charmed strange baryons, $\Xi_c(3055)^0$, $\Xi_c(3055)^+$ and $\Xi_c(3080)^+$ are calculated in a $^3P_0$ model. In the calculation, they are considered as $34$ kinds of $D-wave$ charmed strange baryons. Among these assignments, $\Xi_c(3055)^+$ is very possibly a $J^P={5\over 2}^+$ $\hat\Xi_{c3}^\prime(\frac{5}{2}^+)$ or $\check\Xi_{c3}^2(\frac{5}{2}^+)$. In these two assignments, $\Xi_c(3055)^+$ has the total decay width $\Gamma=10.1$ MeV and $\Gamma=7.6$ MeV, respectively. The predicted ratios $\Gamma(\Xi_c(3055)^+ \to \Lambda D^+ )/ \Gamma(\Xi_c(3055)^+ \to \Sigma_c^{++}K^-)=3.39$. $\Xi_c(3055)^+$ is also very possibly a $J^P={7\over 2}^+$ $\hat\Xi_{c3}^\prime(\frac{7}{2}^+)$ or $\check\Xi_{c3}^2(\frac{7}{2}^+)$. In these two assignments, $\Xi_c(3055)^+$ has the total decay width $\Gamma=9.7$ MeV and $\Gamma=6.3$ MeV, respectively. The predicted ratios $\Gamma(\Xi_c(3055)^+ \to \Lambda D^+ )/ \Gamma(\Xi_c(3055)^+ \to \Sigma_c^{++}K^-)=5.91~\rm{or}~6.04$. As a isospin partner of $\Xi_c(3055)^+$, $\Xi_c(3055)^0$ is also very possibly a $J^P={5\over 2}^+$ $\hat\Xi_{c3}^\prime(\frac{5}{2}^+)$ or $\check\Xi_{c3}^2(\frac{5}{2}^+)$. In these assignments, $\Xi_c(3055)^0$ has the total decay width $\Gamma=10.9$ MeV and $\Gamma=7.0$ MeV. The predicted ratios $\Gamma(\Xi_c(3055)^0 \to \Lambda D^0 )/ \Gamma(\Xi_c(3055)^0 \to \Sigma_c^{+}K^-)=4.24~\rm{or}~4.20$. $\Xi_c(3055)^0$ is also very possibly a $J^P={7\over 2}^+$ $\hat\Xi_{c3}^\prime(\frac{7}{2}^+)$ or $\check\Xi_{c3}^2(\frac{7}{2}^+)$. In these assignments, $\Xi_c(3055)^0$ has the total decay width $\Gamma=10.3$ MeV and $\Gamma=7.1$ MeV. The predicted ratios $\Gamma(\Xi_c(3055)^0 \to \Lambda D^0 )/ \Gamma(\Xi_c(3055)^0 \to \Sigma_c^{+}K^-)=7.47~\rm{or}~7.56$. The results agree well with recent experimental data from Belle. $\Xi_c(3080)^+$ seems impossible to be identified with a D-wave charmed strange baryon.
\end{abstract}

%\keywords{charmed strange baryons, $^3P_0$ model}

\maketitle

\section{Introduction \label{sec:introduction}}
Baryons with one $u$ or $d$, one strange and one charmed quark are identified with $\Xi_c$~\cite{pdg,klempt}. In recent years, more and more highly excited charmed strange baryons have been observed. $\Xi_c(2980)$ and $\Xi_c(3080)$ are first observed by Belle~\cite{Chistov:0606051}, and then confirmed by BaBar~\cite{Aubert:0710.5763}. $\Xi_c(3055)^+$ and $\Xi_c(3123)^+$ are also observed by BaBar~\cite{Aubert:0710.5763}, but only $\Xi_c(3055)^+$ is confirmed by Belle~\cite{Kato:2013ynr}.

Recently, three charmed strange baryons are reported by Belle~\cite{Kato:1605.09013}. $\Xi_c(3055)^0$ is first observed in $\Lambda D^0$ mode, $\Xi_c(3055)^+$ and $\Xi_c(3080)^+$ are first observed in $\Lambda D^+$ mode.
The reported masses of $\Xi_c(3055)^0$, $\Xi_c(3055)^+$ and $\Xi_c(3080)^+$ in Ref.~\cite{Kato:1605.09013} are as follows, $$m_{\Xi_c(3055)^0} = 3059.0 \pm 0.5(stat) \pm 0.6(sys) ~\rm{MeV},$$ $$m_{\Xi_c(3055)^+} = 3055.8 \pm 0.4(stat) \pm 0.2(sys) ~\rm{MeV},$$ $$m_{\Xi_c(3080)^+} = 3079.6 \pm 0.4(stat) \pm 0.1(sys)~\rm{MeV}.$$ The total decay widths obtained from $\Lambda D$ modes are, $$\Gamma_{\Xi_c(3055)^0} = 6.4 \pm 2.1(stat) \pm 1.1(sys)~\rm{MeV},$$ $$\Gamma_{\Xi_c(3055)^+} = 7.0 \pm 1.2(stat) \pm 1.5(sys)~\rm{MeV},$$ $$\Gamma_{\Xi_c(3080)^+} < 6.3~\rm{MeV}.$$

In particular, some ratios of branching fractions are measured, $$\frac{\Gamma_{\Xi_{c}(3055)^{+} \to \Lambda D^{+}}}{\Gamma_{\Xi_{c}(3055)^{+} \to \Sigma_{c}^{++}K^{-}}}=5.09\pm1.01(stat)\pm0.76(sys),$$
$$\frac{\Gamma_{\Xi_{c}(3080)^{+} \to \Lambda D^{+}}}{\Gamma_{\Xi_{c}(3080)^{+} \to \Sigma_{c}^{++}K^{-}}}=1.29\pm0.30(stat)\pm0.15(sys),$$
$$\frac{\Gamma_{\Xi_{c}(3080)^{+} \to \Sigma_{c}^{\ast ++}K^{-} }}{\Gamma_{\Xi_{c}(3080)^{+} \to \Sigma_{c}^{++}K^{-}}}=1.07\pm0.27(stat)\pm0.04(sys).$$

In experiments, $J^P$ quantum numbers have not yet been measured for most of the observed charmed baryons so far. How to identify the observed baryons is an important topic in baryon spectroscopy. The spectroscopy of charmed baryons has been studied in many models (see literature~\cite{klempt} and references therein).

%such as the quark mode~\cite{Capstick:2809 (1986),Roncaglia:1722 (1995),Silvestre-Brac:1 (1996),Ebert:034026 (2005),Roberts:2817 (2008),Garcilazo:961 %(2007),Valcarce:217 (2008)}, heavy quark effective theory (HQET)~\cite{Jenkins:447 (1993)}, QCD sum rules\cite{Bagan:367 (1992),huang,Zhi-Gang Wang:231 %(2008),Zhang:094015 (2008)}, lattice QCD\cite{Bowler:3619 (1996),Mathur:014502 (2002),Lewis:094509 (2001),Chiu:471 (2005)} and et al. In particular,

Hadronic decays of $\Xi_c(3080)^+$ have been studied in a heavy hadron chiral perturbation theory~\cite{cheng}, where $\Xi_c(3080)^+$ is suggested a $J^P={5\over 2}^+$ $\Xi_c$. Hadronic decays of $\Xi_c(3080)^+$ and $\Xi_c(3055)$ have both been studied in a chiral quark model~\cite{zhong}, in which $\Xi_c(3080)^+$ and $\Xi_c(3055)$ are suggested a $J^P={1\over 2}^+$ and a $J^P={3\over 2}^+$ $\Xi_c$, respectively. Obviously, theoretical assignment of $\Xi_c(3080)^+$ is different. Furthermore, these theoretical predictions of relevant ratios of branching fractions are in contradiction with present experimental measurements~\cite{Kato:1605.09013}.

$^3P_0$ model is a phenomenological method to study the OZI-allowed hadronic decays of hadrons. In addition to mesons, it is employed successfully to explain the hadronic decays of baryons~\cite{yaouanc1,Capstick:2809 (1986),Roberts:171 (1992),Capstick:1994 (1993),Capstick:4507 (1994),Capstick:S241 (2000),Chong:094017 (2007)}. In this paper, we will study the hadronic decays of $\Xi_c(3080)^+$ and $\Xi_c(3055)$ in the framework of $^3P_0$ model.

The work is organized as follows. In Sec.II, we give a brief review of the $^3P_0$ model. We present our numerical results in Sec.III. In the last section, we give our conclusions and discussions.

%Charmed strange baryon is a kind of hadron that consists of three quarks, c(charm), s(strange) and a q(up or dowm) in the frame of quark model.
%\keywords{charmed strange baryons, $^3P_0$ model}

%\maketitle

\section{Baryon decay in the $^3P_0$ model \label{Sec: $^3P_0$ model}}
$^3P_0$ model is also known as a Quark Pair Creation (QPC) model. It was first proposed by Micu\cite{micu1969} and developed by Yaouanc et al~\cite{yaouanc1,yaouanc2,yaouanc3}. The model has been subsequently employed and developed to study the OZI-allowed hadronic decays of mesons and baryons by many authors not cited here.

In the model, a pair of quark $q\bar{q}$ is assumed to be created from the vacuum with $J^{PC}=0^{++}$, and then to regroup with the quarks from the initial hadron A to form two daughter hadrons B and C. For meson decays, the created quark regroup with the antiquark of the initial meson, the created antiquark regroup with the quark of the initial meson, and two mesons appear in the final states. For baryon decays, one quark of the initial baryon regroups with the created aitiquark to form a meson, and the rest two quarks regroup with the created quark to form a daughter baryon. The process of a baryon decay is shown in Fig. 1.
\begin{figure}
\begin{center}
\includegraphics[height=2.8cm,angle=0,width=6cm]{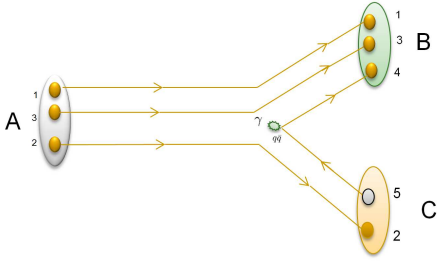}
\caption{Baryon decay process of $A\to B+C$ in the $^3P_0$ model.}
\end{center}
\end{figure}

In the $^3P_0$ model, the hadronic decay width $\Gamma$ of a process $A \to B + C$ is as follows~\cite{yaouanc3},
\begin{eqnarray}
\Gamma  = \pi ^2 \frac{|\vec{p}|}{m_A^2} \frac{1}{2J_A+1}\sum_{M_{J_A}M_{J_B}M_{J_C}} |{\mathcal{M}^{M_{J_A}M_{J_B}M_{J_C}}}|^2.
\end{eqnarray}
In the equation, $\vec{p}$ is the momentum of the daughter baryon in A's center of mass frame,
\begin{eqnarray}
 |\vec{p}|=\frac{{\sqrt {[m_A^2-(m_B-m_C )^2][m_A^2-(m_B+m_C)^2]}}}{{2m_A, }}
\end{eqnarray}
$m_A$ and $J_A$ are the mass and total angular momentum of the initial baryon A, respectively. $m_B$ and $m_C$ are the masses of the final hadrons. $\mathcal{M}^{M_{J_A}M_{J_B}M_{J_C}}$ is the helicity amplitude, which reads~\cite{Chong:094017 (2007)}
\begin{flalign}
 &\mathcal{M}^{M_{J_A } M_{J_B } M_{J_C }}\nonumber \\
 &=-2\gamma\sqrt {8E_A E_B E_C }  \sum_{M_{\rho_A}}\sum_{M_{L_A}}\sum_{M_{\rho_B}}\sum_{M_{L_B}} \sum_{M_{S_1},M_{S_3},M_{S_4},m}  \nonumber\\
 &\langle {J_{l_A} M_{J_{l_A} } S_3 M_{S_3 } }| {J_A M_{J_A } }\rangle \langle {L_{\rho_A} M_{L_{\rho_A} } L_{\lambda_A} M_{L_{\lambda_A} } }| {L_A M_{L_A } }\rangle \nonumber \\
 &\langle L_A M_{L_A } S_{12} M_{S_{12} }|J_{12} M_{J_{12} } \rangle \langle S_1 M_{S_1 } S_2 M_{S_2 }|J_{l_A} M_{J_{l_A} }\rangle \nonumber \\
 &\langle {J_{l_B} M_{J_{l_B} } S_3 M_{S_3 } }| {J_B M_{J_B } }\rangle \langle {L_{\rho_B} M_{L_{\rho_B} } L_{\lambda_B} M_{L_{\lambda_B} } }| {L_B M_{L_B } }\rangle \nonumber \\
 &\langle L_B M_{L_B } S_{14} M_{S_{14} }|J_{14} M_{J_{14} } \rangle \langle S_1 M_{S_1 } S_4 M_{S_4 }|J_{l_B} M_{J_{l_B} }\rangle \nonumber \\
 &\langle {1m;1 - m}|{00} \rangle \langle S_4 M_{S_4 } S_5 M_{S_5 }|1 -m \rangle \nonumber \\
 &\langle L_C M_{L_C } S_C M_{S_C}|J_C M_{J_C} \rangle \langle S_2 M_{S_2 } S_5 M_{S_5 }|S_C M_{S_C} \rangle \nonumber \\
&\times\langle\varphi _B^{1,4,3} \varphi _C^{2,5}|\varphi _A^{1,2,3}\varphi _0^{4,5} \rangle \times I_{M_{L_B } ,M_{L_C } }^{M_{L_A },m} (\vec{p}).
\end{flalign}
In the equation above, $\langle \varphi_B^{1,4,3} \varphi_C^{2,5}|\varphi_A^{1,2,3}\varphi_0^{4,5} \rangle$ is the flavor matrix,
\begin{flalign}
\langle \varphi_B^{1,4,3} \varphi_C^{2,5}|\varphi_A^{1,2,3}\varphi_0^{4,5} \rangle \\
&= f \cdot (-1)^{I_{12}+I_M+I_A+I_3} \nonumber \\
&\times [\frac{1}{2}(2I_M  + 1)(2I_B  + 1)]^{1/2} \nonumber \\
&\times \begin{Bmatrix}
    {I_{12}} & {I_B} & {I_4}\\
    {I_M} & {I_3} & {I_A}\\ \end{Bmatrix}
\end{flalign}
where $f$ takes the value of $(\frac{2}{3})^{1/2}$ or $-(\frac{1}{3})^{1/2}$ according to the isospin $\frac{1}{2}$ or $0$ of the created quarks. $I_{A}$, $I_B$ and $I_M$ represent the isospins of the initial baryon, the final baryon and the final meson. $I_{12}$, $I_{3}$, $I_{4}$ are the isospins of relevant quarks, respectively.

The space integral in Eq.(3) is a little complicated,
\begin{flalign}
I_{M_{L_B } ,M_{L_C } }^{M_{L_A } ,m} (\vec{p})&= \int d \vec{p}_1 d \vec{p}_2 d \vec{p}_3 d \vec{p}_4 d \vec{p}_5 \nonumber \\
&\times\delta ^3 (\vec{p}_1 + \vec{p}_2 + \vec{p}_3 -\vec{p}_A)\delta ^3 (\vec{p}_4+ \vec{p}_5)\nonumber \\
&\times \delta ^3 (\vec{p}_1 + \vec{p}_4 + \vec{p}_3 -\vec{p}_B )\delta ^3 (\vec{p}_2 + \vec{p}_5 -\vec{p}_C) \nonumber \\
& \times\Psi _{B}^* (\vec{p}_1, \vec{p}_4,\vec{p}_3)\Psi _{C}^* (\vec{p}_2 ,\vec{p}_5) \nonumber \\
& \times \Psi _{A} (\vec{p}_1 ,\vec{p}_2 ,\vec{p}_3)y _{1m}\left(\frac{\vec{p_4}-\vec{p}_5}{2}\right).
\end{flalign}

Simple harmonic oscillator (SHO) wave functions are employed as the baryon wave functions\cite{Capstick:2809 (1986),Capstick:1994 (1993),Capstick:4507 (1994)}
\begin{flalign}
\Psi_{A}(\vec{p}_{A})&=N\Psi_{n_{\rho_A} L_{\rho_A} M_{L_{\rho_A}}}(\vec{p}_{\rho_A}) \Psi_{n_{\lambda_A} L_{\lambda_A} M_{L_{\lambda_A}}}(\vec{p}_{\lambda_A}),
\end{flalign}
\begin{flalign}
\Psi_{B}(\vec{p}_{B})&=N\Psi_{n_{\rho_B} L_{\rho_B} M_{L_{\rho_B}}}(\vec{p}_{\rho_B}) \Psi_{n_{\lambda_B} L_{\lambda_B} M_{L_{\lambda_B}}}(\vec{p}_{\lambda_B}),
\end{flalign}
where $N$ represents a normalization coefficient of the total wave function and
\begin{flalign}
\Psi_{nLM_L}(\vec{p})&=\frac{(-1)^n(-i)^L}{\beta^{3/2}}\sqrt{\frac{2n!}{\Gamma(n+L+\frac{3}{2})}}\big(\frac{\vec{p}}{\beta}\big)^L \exp(-\frac{\vec{p}^2}{2\beta^2}) \nonumber\\
&\times L_n^{L+1/2}\big(\frac{\vec{p}^2}{\beta^2}\big)Y_{LM_L}(\Omega_p).
\end{flalign}
$L_n^{L+1/2}\big(\frac{\vec{p}^2}{\beta^2}\big)$ denotes the Laguerre polynomial function, $Y_{LM_L}(\Omega_p)$ is a spherical harmonic function. The relation between the solid harmonica polynomial $y _{LM}(\vec{p})$ and $Y_{LM_L}(\Omega_{\vec{p}})$ is $y _{LM}(\vec{p})=|\vec{p}|^L Y_{LM_L}(\Omega_p)$.

More details of baryon decays in the $^3P_0$ model could be found in Ref.\cite{Capstick:1994 (1993),yaouanc3,Chong:094017 (2007)}. The exact expressions of SHO wave functions of relevant baryons and integrals in momentum space are presented in Appendix A and B, respectively.

\section{Numerical results \label{Sec: Numerical results}}
\subsection{Notations of baryons and relevant parameters}

In the following calculation, notations for the excited baryons are the same as those in Ref.~\cite{Chong:094017 (2007)}. For the hadronic decays of $\Xi_c(3055)$ and $\Xi_c(3080)$, the quantum numbers of initial state baryons and final state baryons are presented in Table I and Table II, respectively.

\begin{table}[t]
\caption{Quantum numbers of initial baryons}
\begin{tabular}{p{0.0cm} p{2.5cm}*{6}{p{0.8cm}}}
   \hline\hline
             & Assignments                                         & $J$                         & $J_l$ & $L_\rho$ & $L_\lambda$ & $L$  & $S_\rho$ \\
   \hline
\label{01}   &$\Xi_{c1}^{' }(\frac{1}{2}^+,\frac{3}{2}^+)$        & $\frac{1}{2}$,$\frac{3}{2}$ &  1    &  0       &   2         &  2   &  1       \\
\label{03}   &$\Xi_{c2}^{' }(\frac{3}{2}^+,\frac{5}{2}^+)$        & $\frac{3}{2}$,$\frac{5}{2}$ &  2    &  0       &   2         &  2   &  1       \\
\label{05}   &$\Xi_{c3}^{' }(\frac{5}{2}^+,\frac{7}{2}^+)$        & $\frac{3}{2}$,$\frac{5}{2}$ &  3    &  0       &   2         &  2   &  1       \\
\label{07}   &$\Xi_{c2}^{  }(\frac{3}{2}^+,\frac{5}{2}^+)$        & $\frac{3}{2}$,$\frac{5}{2}$ &  2    &  0       &   2         &  2   &  0       \\
\label{09}   &$\hat\Xi_{c1}^{' }(\frac{1}{2}^+,\frac{3}{2}^+)$    & $\frac{1}{2}$,$\frac{3}{2}$ &  1    &  2       &   0         &  2   &  1       \\
\label{11}   &$\hat\Xi_{c2}^{' }(\frac{3}{2}^+,\frac{5}{2}^+)$    & $\frac{3}{2}$,$\frac{5}{2}$ &  2    &  2       &   0         &  2   &  1       \\
\label{13}   &$\hat\Xi_{c3}^{' }(\frac{5}{2}^+,\frac{7}{2}^+)$    & $\frac{3}{2}$,$\frac{5}{2}$ &  3    &  2       &   0         &  2   &  1       \\
\label{15}   &$\hat\Xi_{c2}^{ }(\frac{3}{2}^+,\frac{5}{2}^+)$     & $\frac{3}{2}$,$\frac{5}{2}$ &  2    &  2       &   0         &  2   &  0       \\
\label{17}   &$\check\Xi_{c0}^{'0}(\frac{1}{2}^+)$                & $\frac{1}{2}$               &  0    &  1       &   1         &  0   &  0       \\
\label{18}   &$\check\Xi_{c1}^{'1}(\frac{1}{2}^+,\frac{3}{2}^+)$  & $\frac{1}{2}$,$\frac{3}{2}$ &  1    &  1       &   1         &  1   &  0       \\
\label{20}   &$\check\Xi_{c2}^{'2}(\frac{3}{2}^+,\frac{5}{2}^+)$  & $\frac{3}{2}$,$\frac{5}{2}$ &  2    &  1       &   1         &  2   &  0       \\
\label{22}   &$\check\Xi_{c1}^{\ 0}(\frac{1}{2}^+,\frac{3}{2}^+)$ & $\frac{1}{2}$,$\frac{3}{2}$ &  1    &  1       &   1         &  0   &  1       \\
\label{24}   &$\check\Xi_{c0}^{\ 1}(\frac{1}{2}^+)$               & $\frac{1}{2}$               &  0    &  1       &   1         &  1   &  1       \\
\label{25}   &$\check\Xi_{c1}^{\ 1}(\frac{1}{2}^+,\frac{3}{2}^+)$ & $\frac{1}{2}$,$\frac{3}{2}$ &  1    &  1       &   1         &  1   &  1       \\
\label{27}   &$\check\Xi_{c2}^{\ 1}(\frac{3}{2}^+,\frac{5}{2}^+)$ & $\frac{3}{2}$,$\frac{5}{2}$ &  2    &  1       &   1         &  1   &  1       \\
\label{29}   &$\check\Xi_{c1}^{\ 2}(\frac{1}{2}^+,\frac{3}{2}^+)$ & $\frac{1}{2}$,$\frac{3}{2}$ &  1    &  1       &   1         &  2   &  1       \\
\label{31}   &$\check\Xi_{c2}^{\ 2}(\frac{3}{2}^+,\frac{5}{2}^+)$ & $\frac{3}{2}$,$\frac{5}{2}$ &  2    &  1       &   1         &  2   &  1       \\
\label{33}   &$\check\Xi_{c3}^{\ 2}(\frac{5}{2}^+,\frac{7}{2}^+)$ & $\frac{5}{2}$,$\frac{7}{2}$ &  3    &  1       &   1         &  2   &  1       \\
   \hline\hline
\end{tabular}
\label{table1}
\end{table}

\begin{table}[t]
\caption{Quantum numbers of baryons in the final states }
\begin{tabular}{p{0.0cm} p{2.0cm}*{6} {p{0.8cm}}}
   \hline\hline
   &$            $       & $J$           & $J_l$ & $L_\rho$ & $L_\lambda$ & $L$  & $S_\rho$ \\
   \hline
   &$\Xi_c^{0(+)}$       & $\frac{1}{2}$ &  0    &  0       &   0         &  0   &  0  \\
   &$\Xi_c^{'0(+)}$      & $\frac{1}{2}$ &  1    &  0       &   0         &  0   &  1  \\
   &$\Xi_c^{*0(+)}$      & $\frac{3}{2}$ &  1    &  0       &   0         &  0   &  1  \\
   &$\Sigma_c^{0(+,++)}$ & $\frac{1}{2}$ &  1    &  0       &   0         &  0   &  1  \\
   &$\Sigma_c^{*0(+,++)}$& $\frac{3}{2}$ &  1    &  0       &   0         &  0   &  1  \\
   &$\Lambda_c^+$        & $\frac{1}{2}$ &  0    &  0       &   0         &  0   &  0  \\
  % &$\Sigma^{0(+)}$      & $\frac{1}{2}$ &  1    &  0       &   0         &  0   &  1  \\
   &$\Lambda$            & $\frac{1}{2}$ &  0    &  0       &   0         &  0   &  0  \\
   \hline\hline
\end{tabular}
\label{table1}
\end{table}

In these two tables, $L_\rho$ denotes the orbital angular momentum between the two light quarks, $L_\lambda$ denotes the orbital angular momentum between the charm quark and the two light quark system, $L$ is the total orbital angular momentum of $L_\rho$ and $L_\lambda$. $S_\rho$ denotes the total spin of the two light quarks, $J_l$ is total angular momentum of $L$ and $S_\rho$. $J$ is the total angular momentum of the baryons.

$\Xi_c(3080)$ is suggested a D-wave baryon in Ref.~\cite{Chong:094017 (2007)}. $\Xi_c(3055)$ is also suggested a D-wave baryon in Ref.~\cite{zhong}, where $\Xi_c(3080)$ is suggested a radially excited $2S$ baryon. In Ref.~\cite{chen}, both $\Xi_c(3080)$ and $\Xi_c(3055)$ are suggested the D-wave baryons. Therefore, these two states are regarded as the D-wave baryons in our calculation. For the complicated internal combinations, $34$ possible assignments of the D-wave baryons are taken into account in Table. I. The hat and the check are also used to denote the assignments with $L_\rho=2$ and $L_\rho=1$, respectively. The superscript $L$ is adopted to denote the different total angular momentum in $\check\Xi_{cJ_l}^{\ L}$.

Masses of relevant mesons and baryons involved in the calculation are listed in Table III~\cite{pdg}.
\begin{table}[t]
\caption{Masses of involved mesons and baryons in the decays~\cite{pdg}}
\begin{tabular}{p{0.0cm} p{2.0cm}p{2.0cm}|p{2.0cm}p{2.0cm}}
   \hline\hline
   &State              &mass (MeV)  & State          &mass (MeV)\\
   \hline
   &$\pi^{\pm}      $ &139.570    &$\Xi_c^+    $ &2286.46  \\
   &$\pi^{0}        $ &134.977    &$\Xi_c^0    $ &2467.93  \\
   &$K^{\pm}        $ &493.677    &$\Xi_c^{'+} $ &2575.70  \\
   &$K^{0}          $ &497.611    &$\Xi_c^{'0} $ &2577.90  \\
   &$D^{\pm}        $ &1869.61    &$\Xi_c^{*+} $ &2645.90  \\
   &$D^{0}          $ &1864.84    &$\Xi_c^{*0} $ &2645.90  \\
   &$\Lambda        $ &1115.68    &              &         \\
   %&$\Sigma^{+}     $ &1189.37    &$           $ &         \\ &$\Sigma^{0} $ &1192.64
   \hline\hline
\end{tabular}
\label{table1}
\end{table}

Parameters $\gamma$ and $\beta$ are taken as those in Refs.~\cite{Chong:094017 (2007),Blundell:3700 (1996)}. $\gamma=13.4$. $\beta$ is chosen as $476$ MeV for meson $\pi$ and $K$, and $\beta$ is chosen as $435$ MeV for $D$ meson. For baryons, a universal value $\beta=600$ MeV is employed. The mass of $\Xi_c(3055)^0$, $\Xi_c(3055)^+$ and $\Xi_c(3080)^+$ is taken as $3059.0$ MeV, $3055.8$ MeV and $3079.6$ MeV, respectively.

\subsection{Decays of $\Xi_c(3055)$ and $\Xi_c(3080)$}
In general, $u\bar u$, $d\bar d$ and $s\bar s$ could be created from the vacuum. However, there exists no experimental signal for the decay mode with a $s\bar s$ creation. Therefore, the OZI-allowed channels are all assumed from the $u\bar u$ and $d\bar d$ pairs creation in our study. Possible decay modes and corresponding hadronic decay widths of $\Xi_c(3055)^+$, $\Xi_c(3055)^0$, $\Xi_c(3080)^+$ as D-wave state baryons in different assignments are computed and presented in Table IV, Table V and Table VI, respectively. The vanish modes in these three tables indicate forbidden channels or channels with very small decay width. Some ratios of branching fractions related to experiments are particularly given in these tables.

\begin{center}
  \begin{table*}[htbp]
     \caption{Decay widths (MeV) of $\Xi_c(3055)^+$ in different assignments. $\mathcal{B}1$ represents the ratio of branching fractions $\Gamma(\Xi_c(3055)^+ \to \Lambda D^+ )/ \Gamma(\Xi_c(3055)^+ \to \Sigma_c^{++}K^-)$.}
     \begin{tabular}{c ccccccccccccc} \hline \hline
       Assignment            &$\Xi_c^0\pi^+    $ &$\Xi_c^{'0}\pi^+ $ &$\Xi_c^{*0}\pi^+$ &$\Xi_c^{+}\pi^0 $  &$\Xi_c^{'+}\pi^0 $ &$\Xi_c^{*+}\pi^0 $ &$\Sigma_c^{+}K^0$
                             &$\Sigma_c^{*+}K^0$ &$\Lambda_c^{+}K^0$ &$\Sigma_c^{++}K^-$&$\Sigma_c^{*++}K^-$&$\Lambda D^+$      &$\mathcal{B}1 $ \\
     \hline
\label{01}$\Xi_{c1}^{' }(\frac{1}{2}^+)$        &13.0  &2.5                  &0.8    &13.2    &2.6    &0.8    &1.5     &0.2     &16.0     &1.5                  &0.2     &7.0                   &4.69   \\  %01
\label{02}$\Xi_{c1}^{' }(\frac{3}{2}^+)$        &13.0  &0.6                  &1.9    &13.2    &0.6    &2.0    &0.4     &0.4     &16.0     &0.4                  &0.5     &7.0                   &18.80 \\  %02
\label{03}$\Xi_{c2}^{' }(\frac{3}{2}^+)$        &0.0   &5.7                  &0.7    &0.0     &5.8    &0.7    &3.3     &0.2     &0.0      &3.4                  &0.2     &0.0                   &0.00   \\  %03
\label{04}$\Xi_{c2}^{' }(\frac{5}{2}^+)$        &0.0   &$3.3\times10^{-2}$   &4.2    &0.0     &0.0    &4.3    &0.0     &0.9     &0.0      &$5.1\times10^{-3}$   &1.0     &0.0                   &0.00   \\  %04
\label{05}$\Xi_{c3}^{' }(\frac{5}{2}^+)$        &0.3   &$3.8\times10^{-2}$   &0.0    &0.3     &0.0    &0.0    &0.0     &0.0     &0.3      &$5.8\times10^{-3}$   &0.0     &$1.9\times10^{-2}$    &3.30   \\  %05
\label{06}$\Xi_{c3}^{' }(\frac{7}{2}^+)$        &0.3   &$2.2\times10^{-2}$   &0.0    &0.3     &0.0    &0.0    &0.0     &0.0     &0.3      &$3.3\times10^{-3}$   &0.0     &$1.9\times10^{-2}$    &5.88   \\  %06
\label{07}$\Xi_{c2}^{  }(\frac{3}{2}^+)$        &0.0   &3.8                  &0.5    &0.0     &3.9    &0.5    &2.2     &0.1     &0.0      &2.2                  &0.1     &0.0                   &0.00   \\  %07
\label{08}$\Xi_{c2}^{  }(\frac{5}{2}^+)$        &0.0   &$5.0\times10^{-2}$   &2.8    &0.0     &0.1    &2.8    &0.0     &0.6     &0.0      &$7.6\times10^{-3}$   &0.7     &0.0                   &0.00   \\  %08
\label{09}$\hat\Xi_{c1}^{' }(\frac{1}{2}^+)$    &115.1 &22.5                 &7.0    &116.8   &22.9   &7.0    &13.0    &1.6     &141.8    &13.4                 &1.7     &62.6                  &4.68   \\  %09
\label{10}$\hat\Xi_{c1}^{' }(\frac{3}{2}^+)$    &115.1 &5.6                  &17.4   &116.8   &5.7    &17.5   &3.3     &3.9     &141.8    &3.3                  &4.3     &62.6                  &18.73  \\  %10
\label{11}$\hat\Xi_{c2}^{' }(\frac{3}{2}^+)$    &0.0   &50.6                 &6.4    &0.0     &51.6   &6.5    &29.3    &1.4     &0.0      &30.1                 &1.6     &0.0                   &0.00   \\  %11
\label{12}$\hat\Xi_{c2}^{' }(\frac{5}{2}^+)$    &0.0   &0.3                  &37.6   &0.0     &0.3    &38.0   &0.0     &8.5     &0.0      &$5.1\times10^{-2}$   &9.3     &0.0                   &0.00   \\  %12
\label{13}$\hat\Xi_{c3}^{' }(\frac{5}{2}^+)$    &2.7   &0.4                  &0.1    &2.8     &0.4    &0.1    &0.1     &0.0     &3.2      &$5.8\times10^{-2}$   &0.0     &0.2                   &3.39   \\  %13
\label{14}$\hat\Xi_{c3}^{' }(\frac{7}{2}^+)$    &2.7   &0.2                  &0.2    &2.8     &0.2    &0.2    &0.0     &0.0     &3.2      &$3.3\times10^{-2}$   &0.0     &0.2                   &5.91   \\  %14
\label{15}$\hat\Xi_{c2}^{ }(\frac{3}{2}^+)$     &0.0   &33.7                 &4.4    &0.0     &34.4   &4.5    &19.5    &0.9     &0.0      &20.0                 &1.0     &0.0                   &0.00   \\  %15
\label{16}$\hat\Xi_{c2}^{ }(\frac{5}{2}^+)$     &0.0   &0.5                  &25.1   &0.0     &0.5    &25.4   &0.1     &5.6     &0.0      &$7.6\times10^{-2}$   &6.2     &0.0                   &0.00   \\  %16
\label{17}$\check\Xi_{c0}^{'0}(\frac{1}{2}^+)$  &0.0   &43.7                 &54.5   &0.0     &44.6   &55.0   &25.6    &12.5    &0.0      &26.3                 &13.7    &0.0                   &0.00   \\  %17
\label{18}$\check\Xi_{c1}^{'1}(\frac{1}{2}^+)$  &0.0   &0.0                  &0.0    &0.0     &0.0    &0.0    &0.0     &0.0     &0.0      &0.0                  &0.0     &0.0                   & --    \\  %18
\label{19}$\check\Xi_{c1}^{'1}(\frac{3}{2}^+)$  &0.0   &0.0                  &0.0    &0.0     &0.0    &0.0    &0.0     &0.0     &0.0      &0.0                  &0.0     &0.0                   & --    \\  %19
\label{20}$\check\Xi_{c2}^{'2}(\frac{3}{2}^+)$  &0.0   &22.5                 &3.0    &0.0     &22.9   &2.9    &13.0    &0.6     &0.0      &13.4                 &0.7     &0.0                   &0.00   \\  %20
\label{21}$\check\Xi_{c2}^{'2}(\frac{5}{2}^+)$  &0.0   &0.3                  &16.8   &0.0     &0.3    &16.9   &0.0     &3.8     &0.0      &$5.1\times10^{-2}$   &4.1     &0.0                   &0.00   \\  %21
\label{22}$\check\Xi_{c1}^{\ 0}(\frac{1}{2}^+)$ &73.3  &58.3                 &18.2   &74.4    &59.4   &18.3   &34.2    &4.2     &90.4     &35.1                 &4.6     &49.5                  &1.41   \\  %22
\label{23}$\check\Xi_{c1}^{\ 0}(\frac{3}{2}^+)$ &73.3  &14.6                 &45.4   &74.4    &14.9   &45.9   &8.5     &10.4    &90.4     &8.8                  &11.4    &49.5                  &5.64   \\  %23
\label{24}$\check\Xi_{c0}^{\ 1}(\frac{1}{2}^+)$ &0.0   &0.0                  &0.0    &0.0     &0.0    &0.0    &0.0     &0.0     &0.0      &0.0                  &0.0     &0.0                   & --    \\  %24
\label{25}$\check\Xi_{c1}^{\ 1}(\frac{1}{2}^+)$ &0.0   &0.0                  &0.0    &0.0     &0.0    &0.0    &0.0     &0.0     &0.0      &0.0                  &0.0     &0.0                   & --    \\  %25
\label{26}$\check\Xi_{c1}^{\ 1}(\frac{3}{2}^+)$ &0.0   &0.0                  &0.0    &0.0     &0.0    &0.0    &0.0     &0.0     &0.0      &0.0                  &0.0     &0.0                   & --    \\  %26
\label{27}$\check\Xi_{c2}^{\ 1}(\frac{3}{2}^+)$ &0.0   &0.0                  &0.0    &0.0     &0.0    &0.0    &0.0     &0.0     &0.0      &0.0                  &0.0     &0.0                   & --    \\  %27
\label{28}$\check\Xi_{c2}^{\ 1}(\frac{5}{2}^+)$ &0.0   &0.0                  &0.0    &0.0     &0.0    &0.0    &0.0     &0.0     &0.0      &0.0                  &0.0     &0.0                   & --    \\  %28
\label{29}$\check\Xi_{c1}^{\ 2}(\frac{1}{2}^+)$ &76.7  &15.0                 &4.6    &77.8    &15.3   &4.7    &8.7     &1.0     &94.5     &8.9                  &1.1     &41.7                  &4.68   \\  %29
\label{30}$\check\Xi_{c1}^{\ 2}(\frac{3}{2}^+)$ &76.7  &3.7                  &11.6   &77.8    &3.8    &11.7   &2.2     &2.6     &94.5     &2.2                  &3.9     &41.7                  &18.73  \\  %30
\label{31}$\check\Xi_{c2}^{\ 2}(\frac{3}{2}^+)$ &0.0   &33.7                 &4.3    &0.0     &34.4   &4.3    &19.5    &0.9     &0.0      &20.0                 &1.0     &0.0                   &0.00   \\  %31
\label{32}$\check\Xi_{c2}^{\ 2}(\frac{5}{2}^+)$ &0.0   &0.2                  &25.1   &0.0     &0.2    &25.3   &0.0     &5.6     &0.0      &$3.4\times10^{-2}$   &6.2     &0.0                   &0.00   \\  %32
\label{33}$\check\Xi_{c3}^{\ 2}(\frac{5}{2}^+)$ &1.8   &0.3                  &1.0    &1.9     &0.3    &0.1    &0.0     &0.0     &2.1      &$3.9\times10^{-2}$   &0.0     &0.1                   &3.39   \\  %33
\label{34}$\check\Xi_{c3}^{\ 2}(\frac{7}{2}^+)$ &1.8   &0.1                  &0.1    &1.9     &0.1    &0.1    &0.0     &0.0     &2.1      &$2.2\times10^{-2}$   &0.0     &0.1                   &6.04   \\  %34
    \hline\hline
    \end{tabular}
    \label{summary_simfit}
  \end{table*}
\end{center}

\begin{center}
  \begin{table*}[htbp]
     \caption{Decay widths (MeV) of $\Xi_c(3055)^0$ in different assignments. $\mathcal{B}2$ represents the ratio of branching fractions $\Gamma(\Xi_c(3055)^0 \to \Lambda D^0 )/ \Gamma(\Xi_c(3055)^0 \to \Sigma_c^{+}K^-)$. }
     \begin{tabular}{c ccccccccccccc} \hline \hline
       Assignment            &$\Xi_c^0\pi^0    $ &$\Xi_c^{'0}\pi^0 $ &$\Xi_c^{*0}\pi^0$ &$\Xi_c^{+}\pi^- $  &$\Xi_c^{'+}\pi^- $ &$\Xi_c^{*+}\pi^- $ &$\Sigma_c^{0}K^0$
                             &$\Sigma_c^{*0}K^0$ &$\Lambda_c^{+}K^-$ &$\Sigma_c^{+}K^-$ &$\Sigma_c^{*+}K^-$ &$\Lambda D^0$      &$\mathcal{B}2 $ \\
     \hline
\label{01}$\Xi_{c1}^{' }(\frac{1}{2}^+)$        &13.2  &2.6                  &0.8    &13.3    &2.6    &0.8    &1.5     &0.2     &16.4     &1.6                  &0.2     &8.0               &5.08   \\  %01
\label{02}$\Xi_{c1}^{' }(\frac{3}{2}^+)$        &13.2  &0.6                  &2.0    &13.3    &0.7    &2.0    &0.4     &0.5     &16.4     &0.4                  &0.6     &8.0               &20.4  \\  %02
\label{03}$\Xi_{c2}^{' }(\frac{3}{2}^+)$        &0.0   &5.8                  &0.7    &0.0     &5.9    &0.7    &3.4     &0.2     &0.0      &3.6                  &0.2     &0.0               &0.00   \\  %03
\label{04}$\Xi_{c2}^{' }(\frac{5}{2}^+)$        &0.0   &$3.5\times10^{-2}$   &4.4    &0.0     &0.0    &4.3    &0.0     &1.0     &0.0      &$6.0\times10^{-3}$   &0.2     &$2.8\times10^{-2}$&0.00   \\  %04
\label{05}$\Xi_{c3}^{' }(\frac{5}{2}^+)$        &0.3   &$4.0\times10^{-2}$   &0.0    &0.3     &0.0    &0.0    &0.0     &0.0     &0.4      &$7.0\times10^{-3}$   &0.0     &$2.8\times10^{-2}$&4.00   \\  %05
\label{06}$\Xi_{c3}^{' }(\frac{7}{2}^+)$        &0.3   &$2.2\times10^{-2}$   &0.0    &0.3     &0.0    &0.0    &0.0     &0.0     &0.4      &$4.0\times10^{-3}$   &0.0     &0.0               &7.00   \\  %06
\label{07}$\Xi_{c2}^{  }(\frac{3}{2}^+)$        &0.0   &3.9                  &0.5    &0.0     &3.9    &0.5    &2.2     &0.1     &0.0      &2.4                  &0.1     &0.0               &0.00   \\  %07
\label{08}$\Xi_{c2}^{  }(\frac{5}{2}^+)$        &0.0   &$5.3\times10^{-2}$   &2.9    &0.0     &0.1    &2.9    &0.0     &0.7     &0.0      &$9.0\times10^{-3}$   &0.8     &0.0               &0.00   \\  %08
\label{09}$\hat\Xi_{c1}^{' }(\frac{1}{2}^+)$    &117.0 &23.1                 &7.2    &117.9   &23.2   &7.1    &13.4    &1.7     &145.1    &14.2                 &2.0     &71.7              &5.07   \\  %09
\label{10}$\hat\Xi_{c1}^{' }(\frac{3}{2}^+)$    &117.0 &5.8                  &18.0   &117.9   &5.8    &17.8   &3.4     &4.3     &145.1    &3.5                  &4.9     &71.7              &20.27  \\  %10
\label{11}$\hat\Xi_{c2}^{' }(\frac{3}{2}^+)$    &0.0   &51.9                 &6.7    &0.0     &52.3   &6.6    &30.2    &1.5     &0.0      &31.8                 &1.8     &0.0               &0.00   \\  %11
\label{12}$\hat\Xi_{c2}^{' }(\frac{5}{2}^+)$    &0.0   &0.4                  &39.0   &0.0     &0.4    &38.6   &0.0     &9.2     &0.0      &$5.9\times10^{-2}$   &10.7    &0.0               &0.00   \\  %12
\label{13}$\hat\Xi_{c3}^{' }(\frac{5}{2}^+)$    &2.8   &0.4                  &0.2    &2.9     &0.4    &0.2    &0.1     &0.0     &3.5      &$6.7\times10^{-2}$   &0.0     &0.3               &4.24   \\  %13
\label{14}$\hat\Xi_{c3}^{' }(\frac{7}{2}^+)$    &2.8   &0.2                  &0.2    &2.9     &0.2    &0.2    &0.0     &0.0     &3.5      &$3.8\times10^{-2}$   &0.0     &0.3               &7.47   \\  %14
\label{15}$\hat\Xi_{c2}^{ }(\frac{3}{2}^+)$     &0.0   &34.6                 &4.6    &0.0     &34.9   &4.6    &20.2    &1.0     &0.0      &21.2                 &1.2     &0.0               &0.00   \\  %15
\label{16}$\hat\Xi_{c2}^{ }(\frac{5}{2}^+)$     &0.0   &0.5                  &26.1   &0.0     &0.5    &25.8   &0.1     &6.1     &0.0      &$8.8\times10^{-2}$   &7.1     &0.0               &0.00   \\  %16
\label{17}$\check\Xi_{c0}^{'0}(\frac{1}{2}^+)$  &0.0   &44.8                 &56.5   &0.0     &45.2   &55.9   &26.5    &13.5    &0.0      &27.9                 &15.7    &0.0               &0.00   \\  %17
\label{18}$\check\Xi_{c1}^{'1}(\frac{1}{2}^+)$  &0.0   &0.0                  &0.0    &0.0     &0.0    &0.0    &0.0     &0.0     &0.0      &0.0                  &0.0     &0.0               & --    \\  %18
\label{19}$\check\Xi_{c1}^{'1}(\frac{3}{2}^+)$  &0.0   &0.0                  &0.0    &0.0     &0.0    &0.0    &0.0     &0.0     &0.0      &0.0                  &0.0     &0.0               & --    \\  %19
\label{20}$\check\Xi_{c2}^{'2}(\frac{3}{2}^+)$  &0.0   &23.1                 &3.1    &0.0     &23.2   &3.0    &13.4    &0.7     &0.0      &14.2                 &0.8     &0.0               &0.00   \\  %20
\label{21}$\check\Xi_{c2}^{'2}(\frac{5}{2}^+)$  &0.0   &0.4                  &17.4   &0.0     &0.4    &17.2   &0.1     &4.1     &0.0      &$5.9\times10^{-2}$   &4.8     &0.0               &0.00   \\  %21
\label{22}$\check\Xi_{c1}^{\ 0}(\frac{1}{2}^+)$ &74.5  &59.8                 &18.8   &75.1    &60.2   &18.6   &35.3    &4.5     &92.4     &37.2                 &5.2     &56.7              &1.53   \\  %22
\label{23}$\check\Xi_{c1}^{\ 0}(\frac{3}{2}^+)$ &74.5  &14.9                 &47.1   &75.1    &15.1   &46.6   &8.8     &11.2    &92.4     &9.3                  &13.1    &56.7              &6.10   \\  %23
\label{24}$\check\Xi_{c0}^{\ 1}(\frac{1}{2}^+)$ &0.0   &0.0                  &0.0    &0.0     &0.0    &0.0    &0.0     &0.0     &0.0      &0.0                  &0.0     &0.0               & --    \\  %24
\label{25}$\check\Xi_{c1}^{\ 1}(\frac{1}{2}^+)$ &0.0   &0.0                  &0.0    &0.0     &0.0    &0.0    &0.0     &0.0     &0.0      &0.0                  &0.0     &0.0               & --    \\  %25
\label{26}$\check\Xi_{c1}^{\ 1}(\frac{3}{2}^+)$ &0.0   &0.0                  &0.0    &0.0     &0.0    &0.0    &0.0     &0.0     &0.0      &0.0                  &0.0     &0.0               & --    \\  %26
\label{27}$\check\Xi_{c2}^{\ 1}(\frac{3}{2}^+)$ &0.0   &0.0                  &0.0    &0.0     &0.0    &0.0    &0.0     &0.0     &0.0      &0.0                  &0.0     &0.0               & --    \\  %27
\label{28}$\check\Xi_{c2}^{\ 1}(\frac{5}{2}^+)$ &0.0   &0.0                  &0.0    &0.0     &0.0    &0.0    &0.0     &0.0     &0.0      &0.0                  &0.0     &0.0               & --    \\  %28
\label{29}$\check\Xi_{c1}^{\ 2}(\frac{1}{2}^+)$ &77.9  &15.4                 &4.8    &78.6    &15.5   &4.8    &9.0     &1.1     &96.7     &9.4                  &1.3     &47.8              &5.07   \\  %29
\label{30}$\check\Xi_{c1}^{\ 2}(\frac{3}{2}^+)$ &77.9  &3.8                  &12.0   &78.6    &3.9    &11.9   &2.2     &2.8     &96.7     &2.4                  &3.3     &47.8              &20.27  \\  %30
\label{31}$\check\Xi_{c2}^{\ 2}(\frac{3}{2}^+)$ &0.0   &34.6                 &4.5    &0.0     &34.9   &4.4    &20.2    &1.0     &0.0      &21.2                 &1.2     &0.0               &0.00   \\  %31
\label{32}$\check\Xi_{c2}^{\ 2}(\frac{5}{2}^+)$ &0.0   &0.2                  &26.0   &0.0     &0.2    &25.7   &0.0     &6.1     &0.0      &$3.9\times10^{-2}$   &7.1     &0.0               &0.00   \\  %32
\label{33}$\check\Xi_{c3}^{\ 2}(\frac{5}{2}^+)$ &1.9   &0.3                  &0.1    &1.9     &0.2    &0.1    &0.0     &0.0     &2.3      &$4.5\times10^{-2}$   &0.0     &0.2               &4.20   \\  %33
\label{34}$\check\Xi_{c3}^{\ 2}(\frac{7}{2}^+)$ &1.9   &0.2                  &0.1    &1.9     &0.2    &0.1    &0.2     &0.0     &2.3      &$2.5\times10^{-2}$   &0.0     &0.2               &7.56   \\  %34
    \hline\hline
    \end{tabular}
    \label{summary_simfit}
  \end{table*}
\end{center}

\begin{center}
  \begin{table*}[htbp]
     \caption{Decay widths (MeV) of $\Xi_c(3080)^+$ in different assignments. $\mathcal{B}3$ and $\mathcal{B}4$ represents the ratios of branching fraction $\Gamma(\Xi_c(3080)^+ \to \Lambda D^+ )/ \Gamma(\Xi_c(3080)^+ \to \Sigma_c^{++}K^-)$ and $\Gamma(\Xi_c(3080)^+ \to \Sigma_c^{*++}K^-)/ \Gamma(\Xi_c(3080)^+ \to \Sigma_c^{++}K^-)$ respectively.}
     \begin{tabular}{c cccccccccccccc} \hline \hline
       Assignment            &$\Xi_c^0\pi^+    $ &$\Xi_c^{'0}\pi^+ $ &$\Xi_c^{*0}\pi^+$ &$\Xi_c^{+}\pi^0 $  &$\Xi_c^{'+}\pi^0 $ &$\Xi_c^{*+}\pi^0 $ &$\Sigma_c^{+}K^0$
                             &$\Sigma_c^{*+}K^0$ &$\Lambda_c^{+}K^0$ &$\Sigma_c^{++}K^-$&$\Sigma_c^{*++}K^-$&$\Lambda D^+$      &$\mathcal{B}3 $ &$\mathcal{B}4 $   \\
     \hline
\label{01}$\Xi_{c1}^{' }(\frac{1}{2}^+)$        &14.3  &2.9                  &0.9    &14.5    &3.0    &0.9    &2.0     &0.4     &17.7     &2.0                  &0.4               &10.2              &5.07  &0.18  \\  %01
\label{02}$\Xi_{c1}^{' }(\frac{3}{2}^+)$        &14.3  &0.7                  &2.3    &14.5    &0.7    &2.4    &0.5     &0.9     &17.7     &0.5                  &0.9               &10.2              &20.27 &1.84  \\  %02
\label{03}$\Xi_{c2}^{' }(\frac{3}{2}^+)$        &0.0   &6.6                  &0.9    &0.0     &6.7    &0.9    &4.4     &0.3     &0.0      &3.4                  &0.3               &0.0               &0.00  &0.07  \\  %03
\label{04}$\Xi_{c2}^{' }(\frac{5}{2}^+)$        &0.0   &$4.8\times10^{-2}$   &5.1    &0.0     &0.1    &5.1    &0.0     &1.9     &0.0      &$1.1\times10^{-2}$   &2.0               &0.0               &0.00  &185.56\\  %04
\label{05}$\Xi_{c3}^{' }(\frac{5}{2}^+)$        &0.4   &$5.4\times10^{-2}$   &0.0    &0.4     &0.1    &0.0    &0.0     &0.0     &0.5      &$1.2\times10^{-2}$   &$1.3\times10^{-3}$&$5.3\times10^{-2}$&4.31  &0.10  \\  %05
\label{06}$\Xi_{c3}^{' }(\frac{7}{2}^+)$        &0.4   &$3.1\times10^{-2}$   &0.0    &0.4     &0.0    &0.0    &0.0     &0.0     &0.5      &$6.0\times10^{-3}$   &$1.7\times10^{-3}$&$5.3\times10^{-2}$&8.89  &0.29  \\  %06
\label{07}$\Xi_{c2}^{  }(\frac{3}{2}^+)$        &0.0   &4.4                  &0.6    &0.0     &4.5    &0.6    &3.0     &0.2     &0.0      &3.0                  &0.2               &0.0               &0.00  &0.07  \\  %07
\label{08}$\Xi_{c2}^{  }(\frac{5}{2}^+)$        &0.0   &$7.1\times10^{-2}$   &3.4    &0.0     &0.1    &3.4    &0.0     &1.3     &0.0      &$1.6\times10^{-2}$   &1.3               &0.0               &0.00  &82.47 \\  %08
\label{09}$\hat\Xi_{c1}^{' }(\frac{1}{2}^+)$    &123.4 &25.9                 &8.4    &128.1   &26.4   &8.4    &17.6    &3.1     &156.6    &17.9                 &3.3               &90.6              &5.05  &0.19  \\  %09
\label{10}$\hat\Xi_{c1}^{' }(\frac{3}{2}^+)$    &123.4 &6.5                  &20.9   &128.1   &6.6    &21.1   &4.4     &7.9     &156.6    &4.5                  &8.3               &90.6              &20.20 &1.85  \\  %10
\label{11}$\hat\Xi_{c2}^{' }(\frac{3}{2}^+)$    &0.0   &58.4                 &7.8    &0.0     &59.4   &7.9    &39.6    &2.8     &0.0      &40.4                 &3.0               &0.0               &0.00  &0.07  \\  %11
\label{12}$\hat\Xi_{c2}^{' }(\frac{5}{2}^+)$    &0.0   &0.5                  &45.4   &0.0     &0.5    &45.8   &0.1     &17.0    &0.0      &1.0                  &18.0              &0.0               &0.00  &168.00\\  %12
\label{13}$\hat\Xi_{c3}^{' }(\frac{5}{2}^+)$    &3.6   &0.5                  &0.2    &3.7     &0.6    &0.2    &0.1     &0.0     &4.5      &1.2                  &$1.3\times10^{-2}$&0.5               &4.41  &0.10  \\  %13
\label{14}$\hat\Xi_{c3}^{' }(\frac{7}{2}^+)$    &3.6   &0.3                  &0.3    &3.7     &0.3    &0.3    &0.1     &0.0     &4.5      &$6.9\times10^{-2}$   &$1.7\times10^{-2}$&0.5               &7.81  &0.25  \\  %14
\label{15}$\hat\Xi_{c2}^{ }(\frac{3}{2}^+)$     &0.0   &38.9                 &5.5    &0.0     &39.6   &5.5    &26.4    &1.9     &0.0      &26.9                 &2.0               &0.0               &0.00  &0.08  \\  %15
\label{16}$\hat\Xi_{c2}^{ }(\frac{5}{2}^+)$     &0.0   &0.7                  &30.4   &0.0     &0.7    &30.6   &0.2     &11.3    &0.0      &0.2                  &12.0              &0.0               &0.00  &74.94 \\  %16
\label{17}$\check\Xi_{c0}^{'0}(\frac{1}{2}^+)$  &0.0   &50.2                 &65.5   &0.0     &51.1   &66.1   &34.5    &25.0    &0.0      &35.2                 &26.4              &0.0               &0.00  &0.75  \\  %17
\label{18}$\check\Xi_{c1}^{'1}(\frac{1}{2}^+)$  &0.0   &0.0                  &0.0    &0.0     &0.0    &0.0    &0.0     &0.0     &0.0      &0.0                  &0.0               &0.0               & --   & --   \\  %18
\label{19}$\check\Xi_{c1}^{'1}(\frac{3}{2}^+)$  &0.0   &0.0                  &0.0    &0.0     &0.0    &0.0    &0.0     &0.0     &0.0      &0.0                  &0.0               &0.0               & --   & --   \\  %19
\label{20}$\check\Xi_{c2}^{'2}(\frac{3}{2}^+)$  &0.0   &25.9                 &3.6    &0.0     &26.4   &3.7    &17.6    &1.3     &0.0      &17.9                 &1.3               &0.0               &0.00  &0.08  \\  %20
\label{21}$\check\Xi_{c2}^{'2}(\frac{5}{2}^+)$  &0.0   &0.5                  &20.2   &0.0     &0.5    &20.4   &0.1     &7.6     &0.0      &0.1                  &8.0               &0.0               &0.00  &74.7  \\  %21
\label{22}$\check\Xi_{c1}^{\ 0}(\frac{1}{2}^+)$ &80.2  &67.0                 &21.8   &81.2    &68.1   &22.0   &46.1    &8.3     &99.3     &47.0                 &8.8               &71.6              &1.52  &0.19  \\  %22
\label{23}$\check\Xi_{c1}^{\ 0}(\frac{3}{2}^+)$ &80.2  &16.7                 &54.6   &81.2    &17.0   &55.0   &11.5    &20.8    &99.3     &11.7                 &22.0              &71.6              &6.09  &1.87  \\  %23
\label{24}$\check\Xi_{c0}^{\ 1}(\frac{1}{2}^+)$ &0.0   &0.0                  &0.0    &0.0     &0.0    &0.0    &0.0     &0.0     &0.0      &0.0                  &0.0               &0.0               & --   & --   \\  %24
\label{25}$\check\Xi_{c1}^{\ 1}(\frac{1}{2}^+)$ &0.0   &0.0                  &0.0    &0.0     &0.0    &0.0    &0.0     &0.0     &0.0      &0.0                  &0.0               &0.0               & --   & --   \\  %25
\label{26}$\check\Xi_{c1}^{\ 1}(\frac{3}{2}^+)$ &0.0   &0.0                  &0.0    &0.0     &0.0    &0.0    &0.0     &0.0     &0.0      &0.0                  &0.0               &0.0               & --   & --   \\  %26
\label{27}$\check\Xi_{c2}^{\ 1}(\frac{3}{2}^+)$ &0.0   &0.0                  &0.0    &0.0     &0.0    &0.0    &0.0     &0.0     &0.0      &0.0                  &0.0               &0.0               & --   & --   \\  %27
\label{28}$\check\Xi_{c2}^{\ 1}(\frac{5}{2}^+)$ &0.0   &0.0                  &0.0    &0.0     &0.0    &0.0    &0.0     &0.0     &0.0      &0.0                  &0.0               &0.0               & --   & --   \\  %28
\label{29}$\check\Xi_{c1}^{\ 2}(\frac{1}{2}^+)$ &84.3  &17.3                 &5.6    &85.4    &17.6   &5.6    &11.7    &2.1     &104.4    &12.0                 &2.2               &60.4              &5.05  &0.19  \\  %29
\label{30}$\check\Xi_{c1}^{\ 2}(\frac{3}{2}^+)$ &84.3  &4.3                  &14.0   &85.4    &4.4    &14.1   &2.9     &5.2     &104.4    &3.0                  &5.5               &60.4              &20.19 &1.85  \\  %30
\label{31}$\check\Xi_{c2}^{\ 2}(\frac{3}{2}^+)$ &0.0   &38.9                 &5.2    &0.0     &39.6   &5.3    &26.4    &1.9     &0.0      &26.9                 &2.0               &0.0               &0.00  &0.07  \\  %31
\label{32}$\check\Xi_{c2}^{\ 2}(\frac{5}{2}^+)$ &0.0   &0.3                  &30.2   &0.0     &0.3    &30.5   &0.1     &11.3    &0.0      &$7.2\times10^{-2}$   &12.0              &0.0               &0.00  &167.6 \\  %32
\label{33}$\check\Xi_{c3}^{\ 2}(\frac{5}{2}^+)$ &2.4   &0.4                  &0.1    &2.5     &0.4    &0.2    &0.1     &0.0     &3.0      &$8.2\times10^{-2}$   &$8.5\times10^{-3}$&0.4               &4.39  &0.10  \\  %33
\label{34}$\check\Xi_{c3}^{\ 2}(\frac{7}{2}^+)$ &2.4   &0.2                  &0.2    &2.5     &0.2    &0.2    &0.0     &0.0     &3.0      &$4.6\times10^{-2}$   &$1.2\times10^{-2}$&0.4               &7.82  &0.25  \\  %34
    \hline\hline
    \end{tabular}
    \label{summary_simfit}
  \end{table*}
\end{center}

$\Xi_c(3055)^+$ is observed in $\Lambda D^+$ mode with the total decay width $\Gamma_{\Xi_{c}(3055)^{+}}=(7.0 \pm 1.2 \pm 1.5)$ MeV. Once the observed ratio of branching fraction is taken into account, the results in Table IV indicate that $\Xi_c(3055)^+$ is very possibly a $J^P={5\over 2}^+$ $\hat\Xi_{c3}^\prime(\frac{5}{2}^+)$ or $\check\Xi_{c3}^2(\frac{5}{2}^+)$. In these two assignments, $\Xi_c(3055)^+$ has the total decay width $\Gamma=10.1$ MeV and $\Gamma=7.6$ MeV, respectively. In addition, theoretical prediction of the ratio reads
$${\Gamma(\Xi_c(3055)^+ \to \Lambda D^+ )\over\Gamma(\Xi_c(3055)^+ \to \Sigma_c^{++}K^-)}=3.39.$$

$\Xi_c(3055)^+$ is also very possibly a $J^P={7\over 2}^+$ $\hat\Xi_{c3}^\prime(\frac{7}{2}^+)$ or $\check\Xi_{c3}^2(\frac{7}{2}^+)$. In these two assignments, $\Xi_c(3055)^+$ has the total decay width $\Gamma=9.7$ MeV and $\Gamma=6.3$ MeV, respectively. The predicted ratios read
$${\Gamma(\Xi_c(3055)^+ \to \Lambda D^+ )\over \Gamma(\Xi_c(3055)^+ \to \Sigma_c^{++}K^-)}=5.91~\rm{or}~6.04.$$

Both the total decay width and the ratio of branching fraction agree well with experimental data from Belle. The decay modes $\Xi^0_c\pi^+$, $\Xi^+_c\pi^0$ and $\Lambda^+_cK^0$ are the dominant ones. Measurement of these modes in the future can provide more information on this state.

$\Xi_c(3055)^0$ is observed in $\Lambda D^0$ mode with the total decay width $\Gamma_{\Xi_c(3055)^0}=(6.4 \pm 2.1 \pm 1.1)$ MeV. Once the quark component of $\Xi_c(3055)^+$ and $\Xi_c(3055)^0$ is taken into account, the $\Xi_c(3055)^0$ is a isospin partner of $\Xi_c(3055)^+$. The results in Table V indicate that $\Xi_c(3055)^0$ is also very possibly a $J^P={5\over 2}^+$ $\hat\Xi_{c3}^\prime(\frac{5}{2}^+)$ or $\check\Xi_{c3}^2(\frac{5}{2}^+)$. In these assignments, $\Xi_c(3055)^0$ has the total decay width $\Gamma=10.9$ MeV and $\Gamma=7.0$ MeV, respectively. The predicted ratios read
$${\Gamma(\Xi_c(3055)^0 \to \Lambda D^0 )\over \Gamma(\Xi_c(3055)^0 \to \Sigma_c^{+}K^-)}=4.24~\rm{or}~4.20.$$

$\Xi_c(3055)^0$ is also very possibly a $J^P={7\over 2}^+$ $\hat\Xi_{c3}^\prime(\frac{7}{2}^+)$ or $\check\Xi_{c3}^2(\frac{7}{2}^+)$. In these assignments, $\Xi_c(3055)^0$ has the total decay width $\Gamma=10.3$ MeV and $\Gamma=7.1$ MeV, respectively. The predicted ratios read
$${\Gamma(\Xi_c(3055)^0 \to \Lambda D^0 )\over \Gamma(\Xi_c(3055)^0 \to \Sigma_c^{+}K^-)}=7.47~\rm{or}~7.56.$$

The total decay width agree also with experimental data from Belle. As a isospin partner of $\Xi_c(3055)^+$, the above predicted ratios of $\Xi_c(3055)^0$ are reasonable.

$\Xi_c(3080)^+$ is observed in $\Lambda D^+$ mode with the total decay width $\Gamma_{\Xi_c(3080)^+}<6.3$ MeV. Together with the observed ratios of branching fractions from Belle, our results in Table VI implies that it is impossible to identify $\Xi_c(3080)^+$ with a D-wave charmed strange baryon.

%Table VI, the branching fractions $\Gamma(\Xi_c(3055)^+ \to \Lambda D^+ )/ \Gamma(\Xi_c(3055)^+ \to \Sigma_c^{++}K^-)$, $\Gamma(\Xi_c(3080)^+ \to \Lambda D^+ )/ %\Gamma(\Xi_c(3080)^+ \to \Sigma_c^{++}K^-)$ and $\Gamma(\Xi_c(3080)^+ \to \Sigma_c^{*++}K^-)/ \Gamma(\Xi_c(3080)^+ \to \Sigma_c^{++}K^-)$ are listed as can been %seen. The experimental values of the decay widths of $\Xi_c(3055)^+$ and $\Xi_c(3080)^+$ from the $\Lambda D$ final states are $(7.8 \pm 1.2 \pm 1.5) MeV$ and %$<6.3MeV$. Our results of the relevant ratios of branching fractions are also considerable.

%There are some assignments that can be excluded as their decay widths are all zero, like $\check\Xi_{c1}^{'1}(\frac{1}{2}^+)$, %$\check\Xi_{c1}^{'1}(\frac{3}{2}^+)$, $\check\Xi_{c0}^{\ 1}(\frac{1}{2}^+)$, $\check\Xi_{c1}^{\ 1}(\frac{1}{2}^+)$, $\check\Xi_{c1}^{\ 1}(\frac{3}{2}^+)$, %$\check\Xi_{c2}^{\ 1}(\frac{3}{2}^+)$ and $\check\Xi_{c2}^{\ 1}(\frac{5}{2}^+)$. In the process of calculation, we find some common factors that lead to these %results. There are some symmetries among the Clebsch-Gordan coefficients that can yield zero of Eq.(3), and we get zero for the decay widths finally when using %the quantum numbers of those assignments.

\section{Conclusions and discussions\label{Sec: summary}}

In this work, the hadronic decays of $\Xi_c$ baryons are studied in a $^3P_0$ model. We calculate the decay widths and some ratios of branching fractions of $\Xi_c(3055)^0$, $\Xi_c(3055)^+$, $\Xi_c(3080)^+$ related to recent Belle experiment. In comparison with experiments, we make an identification of these $\Xi_c$ baryons. Our theoretical predictions are consistent with experiments.

$\Xi_c(3055)^+$ is observed in $\Lambda D^+$ mode with the total decay width of $(7.0 \pm 1.2 \pm 1.5)$ MeV. Our study indicates that
$\Xi_c(3055)^+$ is very possibly a $J^P={5\over 2}^+$ $\hat\Xi_{c3}^\prime(\frac{5}{2}^+)$ or $\check\Xi_{c3}^2(\frac{5}{2}^+)$. In these two assignments, $\Xi_c(3055)^+$ has the total decay width $\Gamma=10.1$ MeV and $\Gamma=7.6$ MeV, respectively. In addition, we obtain the ratio of branching fraction
${\Gamma(\Xi_c(3055)^+ \to \Lambda D^+ )\over\Gamma(\Xi_c(3055)^+ \to \Sigma_c^{++}K^-)}=3.39.$

$\Xi_c(3055)^+$ is also very possibly a $J^P={7\over 2}^+$ $\hat\Xi_{c3}^\prime(\frac{7}{2}^+)$ or $\check\Xi_{c3}^2(\frac{7}{2}^+)$. In these two assignments, we obtain the total decay width $\Gamma=9.7$ MeV and $\Gamma=6.3$ MeV, respectively. The ratios of branching fraction
${\Gamma(\Xi_c(3055)^+ \to \Lambda D^+ )\over \Gamma(\Xi_c(3055)^+ \to \Sigma_c^{++}K^-)}=5.91~\rm{or}~6.04$ are also obtained.

These theoretical predictions in the $^3P_0$ model agree well with Belle experiments. As the dominant hadronic decay modes of $\Xi_c(3055)^+$, measurement of the modes $\Xi^0_c\pi^+$, $\Xi^+_c\pi^0$ and $\Lambda^+_cK^0$ in the future can provide more information on this state.

As a isospin partner of $\Xi_c(3055)^+$, $\Xi_c(3055)^0$ is observed in $\Lambda D^0$ mode with the total decay width $\Gamma_{\Xi_c(3055)^0}=(6.4 \pm 2.1 \pm 1.1)$ MeV. Our study indicates that $\Xi_c(3055)^0$ is very possibly a $J^P={5\over 2}^+$ $\hat\Xi_{c3}^\prime(\frac{5}{2}^+)$ or $\check\Xi_{c3}^2(\frac{5}{2}^+)$. In these assignments, $\Xi_c(3055)^0$ has the total decay width $\Gamma=10.9$ MeV and $\Gamma=7.0$ MeV, respectively. We predicted the ratios
${\Gamma(\Xi_c(3055)^0 \to \Lambda D^0 )\over \Gamma(\Xi_c(3055)^0 \to \Sigma_c^{+}K^-)}=4.24~\rm{or}~4.20.$

$\Xi_c(3055)^0$ is also very possibly a $J^P={7\over 2}^+$ $\hat\Xi_{c3}^\prime(\frac{7}{2}^+)$ or $\check\Xi_{c3}^2(\frac{7}{2}^+)$. In these assignments, $\Xi_c(3055)^0$ has the total decay width $\Gamma=10.3$ MeV and $\Gamma=7.1$ MeV, respectively. The predicted ratios read
${\Gamma(\Xi_c(3055)^0 \to \Lambda D^0 )\over \Gamma(\Xi_c(3055)^0 \to \Sigma_c^{+}K^-)}=7.47~\rm{or}~7.56.$

$\Xi_c(3080)^+$ is observed in $\Lambda D^+$ mode with the total decay width $\Gamma_{\Xi_c(3080)^+}<6.3$ MeV. Once theoretical predictions of the total decay width and ratios of branching fractions are compared with the experimental data, our study implies that it is impossible to identify $\Xi_c(3080)^+$ with a D-wave charmed strange baryon.

$^3P_0$ model is a phenomenological model to study OZI-allowed hadronic decays of hadron. As well known, there are always uncertainties in normal phenomenological method. As a crosscheck, it is necessary to study the hadronic decay of $\Xi_c$ in much more other models. Of course, further experimental measurement of some modes and their ratios of branching fraction of these states is required.

\begin{acknowledgments}
The authors thank Dr. Yuji Kato from Belle collaboration very much for pointing out the error in our previous manuscript. This work is supported by National Natural Science Foundation of China under the grants: 11075102 and 11475111. It is also supported by the Innovation Program of Shanghai Municipal Education Commission under the grant No. 13ZZ066.
\end{acknowledgments}

\begin{appendix}

\section{Relevant simple harmonic oscillator wave functions}
For the S-wave charmed strange baryon, the total wave function is

\begin{flalign}
\Psi_{0,0,0;0,0,0}(\vec{p}_{\rho},\vec{p}_{\lambda})&=
-3^{3/4}\cdot\frac{e^{-\frac{\vec{p}_{\rho}^2}{2\beta_\rho^2}-\frac{\vec{p}_{\lambda}^2}{2\beta_\lambda^2}}}
{\pi^{3/2}\beta_\rho^{3/2}\beta_\lambda^{3/2}}.
\end{flalign}

For the D-wave charmed strange baryon, the total wave functions are

\begin{flalign}
\Psi_{0,0,0;0,2,m_\lambda}(\vec{p}_{\rho},\vec{p}_{\lambda})&=
-3^{3/4}\cdot\frac{4e^{-\frac{\vec{p}_{\rho}^2}{2\beta_\rho^2}-\frac{\vec{p}_{\lambda}^2}{2\beta_\lambda^2}}
y_{2,m_\lambda}(\vec{p}_{\lambda})}
{\sqrt{15}\pi \beta_\rho^{3/2}\beta_\lambda^{7/2}},
\end{flalign}

\begin{flalign}
\Psi_{0,2,m_\rho;0,0,0}(\vec{p}_{\rho},\vec{p}_{\lambda})&=
-3^{3/4}\cdot\frac{4e^{-\frac{\vec{p}_{\rho}^2}{2\beta_\rho^2}-\frac{\vec{p}_{\lambda}^2}{2\beta_\lambda^2}}
y_{2,m_\rho}(\vec{p}_{\rho})}
{\sqrt{15}\pi \beta_\rho^{7/2}\beta_\lambda^{3/2}},
\end{flalign}

\begin{flalign}
&\Psi_{0,1,m_\rho;0,1,m_\lambda}(\vec{p}_{\rho},\vec{p}_{\lambda})= \nonumber\\
& -3^{3/4}\cdot\frac{8e^{-\frac{\vec{p}_{\rho}^2}{2\beta_\rho^2}-\frac{\vec{p}_{\lambda}^2}{2\beta_\lambda^2}}
y_{1,m_\rho}(\vec{p}_{\rho})y_{1,m_\lambda}(\vec{p}_{\lambda})}
{3\sqrt{\pi} \beta_\rho^{5/2}\beta_\lambda^{5/2}}.
\end{flalign}
The $y_{l,m_l}(\vec{p})$ is the solid harmonic polynomial.

The meson wave function in our calculation is
\begin{flalign}
\Psi_{0,0,0}(\vec{p}_C)&=
\frac{e^{-\frac{ R^2 (\vec{p}_2^2-\vec{p}_5^2)^2}{8}}}
{\pi^{3/4}}.
\end{flalign}

Here $\vec{p}_{\rho}= \sqrt{\frac{1}{2}}(\vec{p}_{1}-\vec{p}_{2})$ , $\vec{p}_{\rho}= \sqrt{\frac{1}{6}}(\vec{p}_{1}+\vec{p}_{2}-2\vec{p}_{3})$ is obtained in the relative Jacobi coordinates. $\vec{p}_C=\frac{\vec{p}_{2}-\vec{p}_{5}}{2}$, $ R=\frac{1}{\beta_c}$.

\section{Momentum space integrations}

The momentum integration $I_{M_{L_B } ,M_{L_C } }^{M_{L_A },m} (\vec{p})$ can be expressed as $I^{l_{\rho_A},M_{\rho_A},l_{\lambda_A},M_{\lambda_A},m}_{M_{L_B},M_{L_C}}$ for the helicity amplitude.

\begin{widetext}
\begin{flalign}
I^{0,0,0,0,0}_{0,0}=\frac{18 p R^{3/2} \left(16+R^2 \beta _{\lambda }^2+3 R^2 \beta _{\rho }^2\right)}{\pi ^{5/4} \left(12+R^2 \beta _{\lambda }^2+3 R^2 \beta _{\rho }^2\right){}^{5/2}} e^{-\frac{p^2 \left(3 R^2 \beta _{\lambda }^4+3 \beta _{\rho }^2 \left(4+R^2 \beta _{\rho }^2\right)+2 \beta _{\lambda }^2 \left(18+7 R^2 \beta _{\rho }^2\right)\right)}{24 \beta _{\lambda }^2 \beta _{\rho }^2 \left(12+R^2 \beta _{\lambda }^2+3 R^2 \beta _{\rho }^2\right)}}.
\end{flalign}

\begin{flalign}
&I^{0,0,2,0,0}_{0,0}= \frac{9 \sqrt{3}p R^{3/2}\left(4+R^2 \beta _{\lambda }^2+R^2 \beta _{\rho }^2\right)}{2 \pi ^{5/4} \beta _{\lambda }^2\left(12+R^2 \beta _{\lambda }^2+3 R^2 \beta _{\rho }^2\right){}^{9/2}}e^{-\frac{p^2 \left(3 R^2 \beta _{\lambda }^4+3 \beta _{\rho }^2 \left(4+R^2 \beta _{\rho }^2\right)+2 \beta _{\lambda }^2 \left(18+7 R^2 \beta _{\rho }^2\right)\right)}{24 \beta _{\lambda }^2 \beta _{\rho }^2 \left(12+R^2 \beta _{\lambda }^2+3 R^2 \beta _{\rho }^2\right)}} \times  \nonumber \\
&\left[R^2 \left(-16+p^2 R^2\right) \beta _{\lambda }^4+4 \beta _{\lambda }^2 \left(-48+5 p^2 R^2+R^2 \left(-12+p^2 R^2\right) \beta _{\rho }^2\right)+p^2 \left(64+28 R^2 \beta _{\rho }^2+3 R^4 \beta _{\rho }^4\right)\right].
\end{flalign}

\begin{flalign}
I^{0,0,2,1,-1}_{0,0}=I^{0,0,2,-1,1}_{0,0}=\frac{108 p R^{3/2} \left(4+R^2 \beta _{\lambda }^2+R^2 \beta _{\rho }^2\right) }{\pi ^{5/4} \left(12+R^2 \beta _{\lambda }^2+3 R^2 \beta _{\rho }^2\right){}^{7/2}} e^{-\frac{p^2 \left(3 R^2 \beta _{\lambda }^4+3 \beta _{\rho }^2 \left(4+R^2 \beta _{\rho }^2\right)+2 \beta _{\lambda }^2 \left(18+7 R^2 \beta _{\rho }^2\right)\right)}{24 \beta _{\lambda }^2 \beta _{\rho }^2 \left(12+R^2 \beta _{\lambda }^2+3 R^2 \beta _{\rho }^2\right)}}.
\end{flalign}

\begin{flalign}
&I^{2,0,0,0,0}_{0,0}=\frac{3 \sqrt{3}\mathcal{E}p R^{3/2}\left(12+R^2 \beta _{\lambda }^2+5 R^2 \beta _{\rho }^2\right)}{2\pi ^{5/4} \beta _{\rho }^2\left(12+R^2 \beta _{\lambda }^2+3 R^2 \beta _{\rho }^2\right){}^{9/2} } e^{-\frac{p^2 \left(3 R^2 \beta _{\lambda }^4+3 \beta _{\rho }^2 \left(4+R^2 \beta _{\rho }^2\right)+2 \beta _{\lambda }^2 \left(18+7 R^2 \beta _{\rho }^2\right)\right)}{24 \beta _{\lambda }^2 \beta _{\rho }^2 \left(12+R^2 \beta _{\lambda }^2+3 R^2 \beta _{\rho }^2\right)}}\times \nonumber \\
&\left[192 p^2+p^2 R^4 \beta _{\lambda }^4+4 \left(-144+29 p^2 R^2\right) \beta _{\rho }^2+3 R^2 \left(-48+5 p^2 R^2\right) \beta _{\rho }^4+4 R^2 \beta _{\lambda }^2 \left(7 p^2+2 \left(-6+p^2 R^2\right) \beta _{\rho }^2\right)\right].
\end{flalign}

\begin{flalign}
I^{2,1,0,0,-1}_{0,0}=I^{2,-1,0,0,1}_{0,0}=\frac{108 p R^{3/2}\left(12+R^2 \beta _{\lambda }^2+5 R^2 \beta _{\rho }^2\right)}{\pi ^{5/4}\left(12+R^2 \beta _{\lambda }^2+3 R^2 \beta _{\rho }^2\right){}^{7/2} }  e^{-\frac{p^2 \left(3 R^2 \beta _{\lambda }^4+3 \beta _{\rho }^2 \left(4+R^2 \beta _{\rho }^2\right)+2 \beta _{\lambda }^2 \left(18+7 R^2 \beta _{\rho }^2\right)\right)}{24 \beta _{\lambda }^2 \beta _{\rho }^2 \left(12+R^2 \beta _{\lambda }^2+3 R^2 \beta _{\rho }^2\right)}}.
\end{flalign}

\begin{flalign}
&I^{2,0,0,0,0}_{0,0}=-\frac{9 \sqrt{3} p R^{3/2}}{2\pi ^{5/4} \beta _{\lambda }\beta _{\rho } \left(12+R^2 \beta _{\lambda }^2+3 R^2 \beta _{\rho }^2\right){}^{9/2} }  e^{-\frac{p^2 \left(3 R^2 \beta _{\lambda }^4+3 \beta _{\rho }^2 \left(4+R^2 \beta _{\rho }^2\right)+2 \beta _{\lambda }^2 \left(18+7 R^2 \beta _{\rho }^2\right)\right)}{24 \beta _{\lambda }^2 \beta _{\rho }^2 \left(12+R^2 \beta _{\lambda }^2+3 R^2 \beta _{\rho }^2\right)}} \times   \nonumber \\
&[\beta _{\lambda }^6 \left(8 R^4-p^2 R^6+4 R^6 \beta _{\rho }^2\right)+\beta _{\lambda }^4 \left(-32 R^2 \left(-6+p^2 R^2\right)+\left(200 R^4-9 p^2 R^6\right) \beta _{\rho }^2+24 R^6 \beta _{\rho }^4\right)- \nonumber \\
&\left(4+R^2 \beta _{\rho }^2\right) \left(192 p^2+4 \left(-72+29 p^2 R^2\right) \beta _{\rho }^2+3 R^2 \left(-24+5 p^2 R^2\right) \beta _{\rho }^4\right)+  \nonumber \\
&\beta _{\lambda }^2 \left(1152-304 p^2 R^2+\left(1920 R^2-176 p^2 R^4\right) \beta _{\rho }^2-23 R^4 \left(-24+p^2 R^2\right) \beta _{\rho }^4+36 R^6 \beta _{\rho }^6\right)].
\end{flalign}

\begin{flalign}
I^{1,1,1,-1,0}_{0,0}=I^{1,-1,1,1,0}_{0,0}=\frac{18 \sqrt{3} p R^{7/2} \beta _{\lambda } \beta _{\rho } \left(16+R^2 \beta _{\lambda }^2+3 R^2 \beta _{\rho }^2\right)}{\pi ^{5/4}\text{  }\left(12+R^2 \beta _{\lambda }^2+3 R^2 \beta _{\rho }^2\right){}^{7/2}}e^{-\frac{p^2 \left(3 R^2 \beta _{\lambda }^4+3 \beta _{\rho }^2 \left(4+R^2 \beta _{\rho }^2\right)+2 \beta _{\lambda }^2 \left(18+7 R^2 \beta _{\rho }^2\right)\right)}{24 \beta _{\lambda }^2 \beta _{\rho }^2 \left(12+R^2 \beta _{\lambda }^2+3 R^2 \beta _{\rho }^2\right)}}.
\end{flalign}

\begin{flalign}
I^{1,0,1,1,-1}_{0,0}=I^{1,0,1,-1,1}_{0,0}=\frac{36 \sqrt{3} p R^{3/2} \frac{\beta _{\lambda }}{\beta _{\rho }} \left(12+R^2 \beta _{\lambda }^2+5 R^2 \beta _{\rho }^2\right)}{\pi ^{5/4} \left(12+R^2 \beta _{\lambda }^2+3 R^2 \beta _{\rho }^2\right){}^{7/2}} e^{-\frac{p^2 \left(3 R^2 \beta _{\lambda }^4+3 \beta _{\rho }^2 \left(4+R^2 \beta _{\rho }^2\right)+2 \beta _{\lambda }^2 \left(18+7 R^2 \beta _{\rho }^2\right)\right)}{24 \beta _{\lambda }^2 \beta _{\rho }^2 \left(12+R^2 \beta _{\lambda }^2+3 R^2 \beta _{\rho }^2\right)}}.
\end{flalign}

\begin{flalign}
I^{1,1,1,0,-1}_{0,0}=I^{1,-1,1,0,1}_{0,0}=\frac{108 \sqrt{3} p R^{3/2} \frac{\beta _{\rho }}{\beta _{\lambda }}\left(4+R^2 \beta _{\lambda }^2+R^2 \beta _{\rho }^2\right)}{\pi ^{5/4} \left(12+R^2 \beta _{\lambda }^2+3 R^2 \beta _{\rho }^2\right){}^{7/2}} e^{-\frac{p^2 \left(3 R^2 \beta _{\lambda }^4+3 \beta _{\rho }^2 \left(4+R^2 \beta _{\rho }^2\right)+2 \beta _{\lambda }^2 \left(18+7 R^2 \beta _{\rho }^2\right)\right)}{24 \beta _{\lambda }^2 \beta _{\rho }^2 \left(12+R^2 \beta _{\lambda }^2+3 R^2 \beta _{\rho }^2\right)}}.
\end{flalign}

\end{widetext}

\end{appendix}


\begin{thebibliography}{99}
\bibitem{pdg}
K.A. Olive et al. (Particle Data Group Collaboration), Chin. Phys. C{\bf 38}, 090001(2014)
\bibitem{klempt}  % ref 02
E. Klempt and Jean-Marc Richard,  Rev. Mod. Phys. {\bf 82}, 1095 (2010).

\bibitem{Chistov:0606051}  % ref 02
R.~Chistov {\it et al.}  (Belle Collaboration),  Phys.\ Rev.\ Lett.\  {\bf 97}, 162001 (2006).
  %[hep-ex/0606051].
\bibitem{Aubert:0710.5763}  % ref 03
   B.~Aubert {\it et al.}  (BaBar Collaboration),  Phys.\ Rev.\ D {\bf 77}, 012002 (2008).
  %[arXiv:0710.5763 [hep-ex]].
\bibitem{Kato:2013ynr}   %09
  Y.~Kato {\it et al.} (Belle Collaboration),  Phys.\ Rev.\ D {\bf 89}, 052003 (2014).
  %[arXiv:1312.1026 [hep-ex]].
\bibitem{Kato:1605.09013}  % ref 01
Y.~Kato {\it et al.} (Belle Collaboration), Phys.\ Rev.\ D {\bf 94}, 032002 (2016).
%\bibitem{Aubert:2006je}   %04
%  B.~Aubert {\it et al.}  [BaBar Collaboration],  Phys.\ Rev.\ Lett.\  {\bf 97}, 232001 (2006).
  %[hep-ex/0608055].
%\bibitem{Aubert:2006sp}   %05
%  B.~Aubert {\it et al.}  [BaBar Collaboration],  Phys.\ Rev.\ Lett.\  {\bf 98}, 012001 (2007).
  %[hep-ex/0603052].
%\bibitem{Abe:2006rz}   %06
%  K.~Abe {\it et al.}  [Belle Collaboration],  Phys.\ Rev.\ Lett.\  {\bf 98}, 262001 (2007).
  %[hep-ex/0608043].
%\bibitem{Solovieva:2008fw}   %07
%  E.~Solovieva {\it et al.},  Phys.\ Lett.\ B {\bf 672}, 1 (2009).
  %[arXiv:0808.3677 [hep-ex]].
%\bibitem{Lesiak:2008wz}   %08
%  T.~Lesiak {\it et al.}  [Belle Collaboration],  Phys.\ Lett.\ B {\bf 665}, 9 (2008).
  %[arXiv:0802.3968 [hep-ex]].
\bibitem{cheng}
Hai-Yang Cheng and Chun-Khiang Chua, Phys. Rev. D \textbf{75}, 014006 (2007).
\bibitem{zhong}
Lei-Hua Liu, Li-Ye Xiao and Xian-Hui Zhong, Phys. Rev. D \textbf{86}, 034024 (2012).
\bibitem{Capstick:2809 (1986)}   %10
S. Capstick and N. Isgur, Phys. Rev. D \textbf{34}, 2809 (1986).
%\bibitem{Roncaglia:1722 (1995)}   %11
%  R. Roncaglia, D. B. Lichtenberg, and E. Predazzi, Phys. Rev. D \textbf{52}, 1722 (1995).
%\bibitem{Silvestre-Brac:1 (1996)}   %12
%  B. Silvestre-Brac, Few Body Syst \textbf{20}, 1 (1996).
%\bibitem{Ebert:034026 (2005)}   %13
%  D. Ebert, R. N. Faustov, and V. O. Galkin, Phys. Rev. D \textbf{72}, 034026 (2005).
%\bibitem{Roberts:2817 (2008)}   %14
%  W. Roberts and M. Pervin, Int. J. Mod. Phys. \textbf{A23}, 2817 (2008).
%\bibitem{Garcilazo:961 (2007)}   %15
%  H. Garcilazo, J. Vijande, and A. Valcarce, J. Phys. \textbf{G34}, 961 (2007).
%\bibitem{Valcarce:217 (2008)}   %16
%  A. Valcarce, H. Garcilazo, and J. Vijande, Eur. Phys. J. \textbf{A37}, 217 (2008).
%\bibitem{Jenkins:447 (1993)}   %17
%  E. Jenkins, Phys. Lett. \textbf{B315}, 447 (1993).
%\bibitem{Bagan:367 (1992)}   %18
%  E. Bagan, M. Chabab, H. G. Dosch, and S. Narison, Phys. Lett. \textbf{B278}, 367 (1992).
%\bibitem{huang}   %19
%Chao-Shang Huang, Ai-Lin Zhang and Shi-Lin Zhu, Phys. Lett. B \textbf{492}, 288 (2000).
%\bibitem{Zhi-Gang Wang:231 (2008)}   %20
%  Zhi-Gang Wang, Eur. Phys. J. \textbf{C54}, 231 (2008).
%\bibitem{Zhang:094015 (2008)}   %21
%  J.-R. Zhang and M.-Q. Huang, Phys. Rev. D \textbf{78}, 094015 (2008).
%\bibitem{Bowler:3619 (1996)}   %22
%  K. Bowler et al. (UKQCD), Phys. Rev. D \textbf{54}, 3619 (1996).
%\bibitem{Mathur:014502 (2002)}   %23
%  N. Mathur, R. Lewis, and R. Woloshyn, Phys. Rev. D \textbf{66}, 014502 (2002).
%\bibitem{Lewis:094509 (2001)}   %24
%  R. Lewis, N. Mathur, and R. Woloshyn, Phys. Rev. D \textbf{64}, 094509 (2001).
%\bibitem{Chiu:471 (2005)}   %25
%  T.-W. Chiu and T.-H. Hsieh, Nucl. Phys.  \textbf{A755}, 471 (2005).
\bibitem{Roberts:171 (1992)}   %27
  W. Roberts and B. Silvestre-Brac, Few Body Syst. \textbf{11}, 171 (1992).
\bibitem{Capstick:1994 (1993)}   %28
  S. Capstick and W. Roberts, Phys. Rev. D \textbf{47}, 1994 (1993).
\bibitem{Capstick:4507 (1994)}   %29
  S. Capstick and W. Roberts, Phys. Rev. D \textbf{49}, 4570 (1994).
\bibitem{Capstick:S241 (2000)}   %30
  S. Capstick and W. Roberts, Prog. Part. Nucl. Phys. \textbf{45}, S241 (2000).
\bibitem{Chong:094017 (2007)}   %31
  Chong Chen {\it et al.}, Phys. Rev. D \textbf{75}, 094017 (2007).

\bibitem{micu1969}   %32
L. Micu, Nucl. Phys. B {\bf 10}, 521 (1969).
\bibitem{yaouanc1}   %33
A. Le Yaouanc, L. Oliver, O. P$\grave{e}$ne and J.C. Raynal, Phys. Rev. D {\bf 8}, 2223 (1973); {\bf 9}, 1415 (1974); {\bf 11}, 1272 (1975).
\bibitem{yaouanc2}   %34
A. Le Yaouanc, L. Oliver, O. P$\grave{e}$ne and J.C. Raynal, Phys. Lett. {\bf 71}, 397 (1977); {\bf 72}, 57 (1977).
\bibitem{yaouanc3}   %35
A. Le Yaouanc, L. Oliver, O. P$\grave{e}$ne and J.C. Raynal, Hadron Transitions in the Quark Model, Gordon and Breach Science Publishers, New York, 1987.
\bibitem{chen}   %36
Bing Chen, Ke-Wei Wei and Ailin Zhang,  Eur. Phys. J.  {\bf A 51} 82 (2015).
\bibitem{Blundell:3700 (1996)}   %38
H.G. Blundell and S. Godfrey, Phys. Rev. D \textbf{53}, 3700 (1996).




%\bibitem{Durr:2015dna}
%S. Durr \textit{et al.}, Phys. Rev. Lett. \textbf{116}, 172001 (2016).
\end{thebibliography}
\end{document}